\documentclass[preprint2]{aastex}
\usepackage{amsmath,amssymb}

\usepackage{graphicx,color}

\usepackage[colorlinks,breaklinks,bookmarks=false]{hyperref}

\newcommand{\MN}{M_{\text{B}}}
\newcommand{\Mq}{M_{\text{q}}}
\newcommand{\Md}{M_{\text{d}}}
\newcommand{\muB}{\mu_{\text{B}}}
\newcommand{\muq}{\mu_{\text{q}}}
\newcommand{\Nc}{N_{\text{c}}}
\newcommand{\Tc}{T_{\text{c}}}
\newcommand{\calO}{\mathcal{O}}
\newcommand{\gv}{G_{\text{v}}}
\newcommand{\gs}{G_{\text{s}}}
\newcommand{\gd}{G_{\text{d}}}
\newcommand{\MeV}{\;\text{MeV}}
\newcommand{\GeV}{\;\text{GeV}}
\newcommand{\rmfm}{\;\text{fm}}
\newcommand{\nB}{n_{\text{B}}}
\newcommand{\LQCD}{\Lambda_{\text{QCD}}}
\newcommand{\one}{\boldsymbol{1}}
\newcommand{\diag}{\text{diag}}

\newcommand{\feyn}[1]{
  \setbox0=\hbox{\ensuremath{#1}}
  \hbox to\wd0{\hbox to0pt{\hbox to\wd0{\hss/\hss}\hss}\box0}}

\begin{document}
\title{The Quarkyonic Star}

\author{Kenji Fukushima}
\affil{Department of Physics, The University of Tokyo,
       7-3-1 Hongo, Bunkyo-ku, Tokyo 113-0033, Japan}
\email{fuku@nt.phys.s.u-tokyo.ac.jp}

\author{Toru Kojo}
\affil{Department of Physics, University of Illinois at
       Urbana-Champaign, 1110 W.\ Green Street, Urbana,
       Illinois 61801, USA}
\email{torukojo@illinois.edu}

\begin{abstract}
  We discuss theoretical scenarios on crossover between nuclear matter
  (NM) and quark matter (QM).  We classify various possibilities into
  three major scenarios according to the onset of diquark degrees of
  freedom that characterizes color-superconducting (CSC) states.  In
  the conventional scenario NM occurs at the liquid-gas (or
  liquid-vacuum at zero temperature) phase transition and QM occurs
  next, after which CSC eventually appears.  With the effect of strong
  correlation, the BEC-BCS scenario implies that CSC occurs next to NM
  and QM comes last in the BCS regime.  We adopt the quarkyonic
  scenario in which NM, QM, and CSC are theoretically
  indistinguishable and thus these names refer to not distinct states
  but relevant descriptions of the same physical system.  Based on
  this idea we propose a natural scheme to interpolate NM near normal
  nuclear density and CSC with vector coupling at high baryon
  density.  We finally discuss the mass-radius relation of the neutron
  star and constraints on parameters in the proposed scheme.
\end{abstract}

\keywords{Neutron star, Equation of state, Nuclear matter,
 Quark matter, Color superconductor}
\maketitle

\section{Introduction}

Quantum chromodynamics (QCD) is one fundamental theory in the Standard
Model in which non-Abelian gauge fields (gluons) and fermions (quarks)
interact non-perturbatively and QCD-related phenomena are
characterized by an energy scale; $\LQCD \simeq 0.2\GeV$.  The
QCD-vacuum has a rich structure with a variety of condensates and it
dynamically confines any colored excitation.  At finite temperature of
$T \sim \LQCD$ and low baryon density of $\nB\ll \LQCD^3$,
ultra-relativistic nucleus-nucleus collision at Relativistic Heavy-Ion
Collider called RHIC and Large Hadron Collider called LHC,
supported by the lattice-QCD Monte-Carlo calculations,
has revealed
quantitative properties of a novel strongly-correlated state of QCD matter, 
hadronic gas at $T\lesssim\LQCD$ and the quark-gluon plasma (QGP) at
$T\gtrsim\LQCD$.  In contrast, when $T\ll\LQCD$ and $n_B\sim \LQCD^3$,
our knowledge from QCD is severely limited because the first-principle
numerical method based on the importance sampling fails at finite
baryon density.  Therefore, it is important to impose constraints on
theoretical uncertainties from the experimental side such as the
neutron star observation and the beam-energy scan program at RHIC and
future heavy-ion facilities.  The purpose of this paper is to import
the state-of-the-art idea of color deconfinement phenomenon at high
$T$ and high $\nB$, especially a special feature of dense and large-$\Nc$ QCD, into the neutron star phenomenology, which in turn
provides us with useful constraints.

In some limiting cases at low and high baryon densities we have
reasonable understanding of cold QCD matter.  At
$\nB = (1\sim 2)n_0$ (where $n_0 \simeq 0.16\rmfm^{-3}$ is the nuclear
saturation density), on the one hand, one can utilize empirical
knowledge from nuclear physics and well-developed theoretical methods
to analyze nuclear matter (NM) properties
\citep{Weinberg:1990rz,Epelbaum:2008ga}.  Recent developments include
not only the chiral perturbation theory
\citep{Fiorilla:2011sr,Kruger:2013kua} and the chiral effective model
\citep{Lastowiecki:2015mpa,Holt:2014hma} but also the functional renormalization
group \citep{Drews:2014spa}.  On the other hand, at asymptotically
high density $n_B \gtrsim 100n_0$ (or even smaller densities hopefully), perturbative QCD (pQCD)
calculations are validated for the bulk quantities of quark matter
(QM) as seen from the convergence of three-loop perturbative results
\citep{Freedman:1976ub,Kurkela:2009gj,Fraga:2015xha}.  Besides, pQCD works well to
describe the color-superconducting (CSC) states.  In particular the
ground state of QCD in the high-density limit has been identified as
the color-flavor-locked (CFL) state \citep{Alford:1998mk} (see also a
review by \citet{Fukushima:2010bq} and references therein).

The most problematic is the QCD matter study in the intermediate
density region: $2n_0 < \nB < 100 n_0$ or $1.1\GeV < \muB < 3\GeV$ in
terms of the baryon chemical potential.  Neutron stars are unique
cosmic laboratories to access such domains experimentally
\citep{Buballa:2014jta}.  Useful observations include the mass-radius
($M$-$R$) relation, the cooling curve \citep{Shternin:2010qi,Page:2004fy,Page:2005fq,Page:2010aw,Tsuruta:2002ey,Blaschke:1999qx,Blaschke:2004vq}, 
the surface and toroidal magnetic fields \citep{Duncan:1992hi,Cardall:2000bs,Olausen:2013bpa,Kitamoto:2014wca}, 
the gravitational waves from the merger of binary neutron stars \citep{Abbott:2007kv,Hotokezaka:2011dh,Hotokezaka:2012ze},
and so on.  In the present work we
will specifically pay our attention to the mass-radius relation.

The interesting point of the $M$-$R$ relation is that it has a
one-to-one correspondence  to the the QCD equation of state (EoS)
through the Tolman-Oppenheimer-Volkoff (TOV) equation.
The overall size of neutron star radii is determined by the EoS around
$n_B=(1\sim 2)n_0$ \citep{Lattimer:2000nx,Lattimer:2006xb}, while the typical neutron
star masses $M \gtrsim M_{\odot}$ (where $M_\odot$ denotes the solar
mass) are largely correlated with the pressure at the central cores
with $n_B = (2\sim 10)n_0$.  According to the discoveries of PSR
J1614-2230 \citep{Demorest:2010bx} and PSR J0348+0432
\citep{Antoniadis:2013pzd} there should be neutron stars whose mass
exceeds $2M_\odot$, indicating that the QCD EoS must be very stiff at
$n_B>2n_0$ as compared to what was na\"{i}vely considered.

The requirement of stiffness challenges conventional EoS's beyond the
nuclear regime.  In typical hadronic models the strangeness appears at
$\nB=(2\sim 3)n_0$ \citep{Glendenning:1997ak,Glendenning:1998zx,Tsubakihara:2009zb}, 
which significantly softens the EoS (i.e., hyperon
puzzle).  Seminal works
\citep{Nishizaki:2002ih,Vidana:2010ip,Weissenborn:2011ut,Yamamoto:2014jga,Lonardoni:2014bwa}
introduced repulsive interactions of strangeness to circumvent the
hyperon puzzle, and the results imply that the repulsion not only in
two-body force of baryons but also in three- and more-body forces are
necessary to pass the $2M_\odot$ constraint.  
While many-body forces including hyperons have been poorly constrained,
new data on the nucleon-hyperon scattering and hypernuclei at J-PARC
\citep{Nagae:2010zz,Tamura:2010zz,Nakazawa:2010zzb} 
as well as the lattice simulations at the physical quark masses
\citep{Inoue:2011ai,Beane:2012vq}
will provide important clues to resolve the hyperon puzzle.

In general contributions from many-body force of baryons starts growing rapidly
at $\nB\sim 2n_0$ \citep{Hebeler:2010jx}.
From the microscopic point of view strong many-body correlations imply
that baryons should exchange many mesons and quarks at $\nB\gtrsim
2n_0$.  With more and more exchanged quarks, the identity of isolated
baryon should be diminished, and quark degrees of freedom gradually
take over physical degrees of freedom.  Eventually, baryons with the
radii $(0.5\sim 0.8)\rmfm$ begin to overlap with each other at
$\nB\gtrsim (5\sim 10)n_0$, leading to the percolated quark matter
\citep{Baym:1976yu}.  This physics picture contrasts with other
crossover scenarios and fills smoothly in a gap between nuclear and
quark matter with exchanged mesons and quarks.  The matter in the
confinement/deconfinement crossover domain should inherit properties
from both nuclear and quark matter.  The idea of
\textit{quarkyonic matter}, that was first recognized by
\citet{McLerran:2007qj}, is an important step toward correct
understanding of the crossover, as we will closely discuss later.  
We will further push this idea forward concrete implementation to
construct an EoS.

In this paper we delineate the crossover scenario bridged by
quarkyonic matter.  As we emphasize later, the most essential point in
this scenario is that there exists an overlapping region that can be
described in terms of strongly-interacting baryons or
strongly-interacting quarks equally.  For a concrete realization we
utilize an Nambu--Jona-Lasinio (NJL) type model 
including vector and diquark interactions (for a review, \cite{Buballa:2003qv})
which are generally density dependent.  Applying
the quark-hadron duality picture we determine the low-density behavior
of parameters by fitting the Akmal-Pandaripande-Ravenhall (APR)
nuclear EoS \citep{Akmal:1998cf}.  The high-density behavior is
constrained by confronting the EoS with the $2M_\odot$ constraint.
In this way we cast the $M$-$R$ relation onto the quark model
parameters, and we analyze the transitional behavior with increasing
density.  

One robust conclusion from our analysis is that the
interactions should remain large even at $\nB\sim 10n_0$ to account
for the existence of massive neutron stars.  
One might consider that
the asymptotic freedom in QCD would rather validate 
weakly interacting picture just in percolated quark matter, 
since the typical distance among quarks is small. Such intuition, however, does not always work
unless soft gluon contributions are cutoff.
For instance, in high-$T$ QCD matter magnetic gluons remain unscreened
and non-perturbative contributions survive even for asymptotically
high temperature \citep{Hietanen:2008tv}.
Also QCD in two-dimensions as an asymptotic free theory \citep{'tHooft:1974hx}
has dense quark matter in which non-perturbative gluons survive to asymptotically high density,
since in spatially one-dimension screening effects are not enhanced at finite density 
\citep{Schon:2000he,Bringoltz:2009ym,Kojo:2011fh}.
In dense QCD, while the hard-dense-loop (HDL) type calculations at weak coupling predict the electric screening mass
of the QCD coupling constant times quark chemical potential, $\sim g_s \mu_q$, 
actually the soft region in the gluon polarization function is highly dependent on the
non-perturbative physics near the quark Fermi surface and can deviate from the
HDL results at {\it qualitative} level \citep{Rischke:2000cn,Huang:2004am,Fukushima:2005cm,Kojo:2014vja}.
Therefore the significance of the screening effects is still an open question until 
first-principle calculation determines the phase structure.
In this situation it should be useful to construct an argument 
in which non-perturbative gluons survive in percolated quark matter,
as discussed in the quarkyonic scenario.
In this paper we try to address this issue from the neutron star phenomenology,
and use the $2M_\odot$ constraint to discuss why non-perturbative gluons should survive to $n_B\sim 10n_0$.

The treatment of the crossover EoS proposed in the present work has some overlap
with preceding works.
The attempts to directly interpolate pQCD and nuclear EoS have been made 
by \cite{Freedman:1977gz,Fraga:2015xha}.
Also \cite{Alford:2004pf} discussed that an EoS for quark matter with CSC can mimic
the APR EoS. 
In these studies the descriptions of quark matter are based on 
the extrapolation of the pQCD picture. 
On the other hand, recent studies on the crossover scenario
have used more model dependent but direct descriptions to conceptualize
strongly correlated quark matter for $n_B \gtrsim (3\sim5)n_0$, 
then interpolated the resulting quark EoS and nuclear EoS at $n_B \lesssim 2n_0$ 
\citep{Masuda:2012ed,Masuda:2012kf,Masuda:2015kha,Alvarez-Castillo:2013spa,Hell:2014xva,Kojo:2014rca,Kojo:2015fua}.
These studies, however, do not manifestly treat microscopic dynamics 
in the interpolated domain, so one cannot address, for instance, how the strangeness changes in the crossover domain. 
In the present work we will take a more direct
and concrete path to the microscopic description with the interplay
among quarks, diquarks, and baryons.  We particularly put our emphasis
on the role played by diquarks, with which there emerges a natural
classification scheme of crossover scenarios as we see below.

This paper is organized as follows.  In Sec.~\ref{sec:diquark} we
classify the crossover scenarios and discuss the underlying physics.  
In Sec.~\ref{sec:model} we define our model and elucidate
how to treat the model parameters.  In Sec.~\ref{sec:EoS} we construct
an EoS and extract the density dependence of the model parameters.  In
Sec.~\ref{sec:MR} we discuss the resulting $M$-$R$ relations and
finally we make concluding remarks in Sec.~\ref{sec:conclusions}.

\section{Quark degrees of freedom beyond nuclear matter}
\label{sec:diquark}

There is no rigorous order parameter for quark deconfinement as soon
as dynamical quarks are included in the theory, irrespective of
various theoretical efforts \citep{Fukushima:2002bk}.  The Polyakov
loop serves for an only approximate order parameter at high $T$, while
we have no clue about even an approximate order parameter if the
baryon density is high (see also an attempt by
\citet{Dexheimer:2009hi} to introduce a parameter analogous to the
Polyakov loop).  Here, we discuss the general features of
deconfinement crossover not relying on any models.

\subsection{Three scenarios}
\label{sec:three}
There is no established understanding on how quarks can become
dominating beyond the normal nuclear density where nucleons are the
most relevant degrees of freedom.  Here we sort out various
theoretical speculations on QM into three clearly distinct categories.
We particularly pay special attention to the onset of diquark degrees
of freedom introducing a diquark binding energy $B_d$ together with a
baryon binding energy $B_b$.  Then, we can express the baryon mass as
$\MN=3\Mq-B_b$ and the diquark mass as $\Md=2\Mq-B_d$, where $\Mq$
represents the constituent quark mass \citep{pawlowski}.

\begin{itemize}
\item {Conventional scenario (NM $<$ QM $<$ CSC):}
In CSC at weak coupling QM is assumed to overcome NM beyond some
baryon density.  Once we admit it, we can see that the Cooper
instability suggests strong correlation of diquarks in momentum space.
This, however, does not necessarily imply that those diquarks are
localized in space or bound states.  In fact we usually have $B_d<0$
in the BCS-type calculation, leading to the following ordering:
$\MN/3 < \Mq < \Md/2$.  Therefore, there should be an onset for
QM next to the liquid-gas 
transition of NM.\ \ Such an onset
is implemented in quark models or a first-order phase transition
between NM and QM around $\muB\sim 3\Mq$ (with minor in-medium
corrections) is often assumed to construct an EoS.\ \ At higher baryon
density, eventually, diquarks form a condensate in the CSC
states.

\item {BEC-BCS scenario (NM $<$ CSC $<$ QM):}
In the intermediate region of baryon density the weak-coupling study
may be drastically altered.  If the attractive interaction is strong
enough, the Cooper pair could become localized in space and then CSC
is identified as a Bose-Einstein condensate (BEC) of bound
diquarks \citep{Abuki:2001be}, which is called a (quasi-)molecule in
condensed matter physics systems controlled by a Feshbach resonance
\citep{Ohashi:2002}.  Then, we should anticipate a crossover
transition between the BCS and the BEC regimes as the coupling
strength changes \citep{Nishida:2005ds,Kitazawa:2007zs,Sun:2007fc}.  In the BEC
regime where $B_d>0$, the mass hierarchy should be reorganized as:
$\MN/3 < \Md/2 < \Mq$.  This is actually a situation in some
descriptions using quark-meson-diquark models in which diquarks are
considered as physical degrees of freedom \citep{pawlowski}.

\item {Quarkyonic scenario (NM $\sim$ QM $\sim$ CSC):}
It is also a speculative but logical possibility that interacting nucleons are, in
principle, indistinguishable from matter out of quarks.  One heuristic
measure to characterize quark deconfinement is the mobility of hadrons
and quarks \citep{Karsch:1979zt}.  Then, quarks can always hop from
one nucleon to the other via meson exchange.  In this sense the
mobility of quarks is never vanishing as long as nucleons are
interacting.  One can upgrade this hand-waving argument to a more
precise formulation by taking the large-$\Nc$ limit, which makes clear
the difference between the high-$T$ low-$\muB$ situation and the
low-$T$ high-$\muB$ situation.  In the former case mesons become
non-interacting objects in the large-$\Nc$ limit, so that quark
deconfinement is well-defined by percolation of overlapping
wave-functions of mesons.  In the latter case at high baryon density,
on the other hand, the interaction among nucleons is of $\calO(\Nc)$
leading to a large pressure of $\calO(\Nc)$ that is comparable to the
pressure of quark matter \citep{McLerran:2007qj}.  This is a natural
consequence from the fact that nucleon interactions are induced by
quark exchange, and thus the pressure of nuclear matter should be
sensitive to quark degrees of freedom.  This important observation on
the pressure of $\calO(\Nc)$ of both nuclear and quark matter opens a
third scenario that there should be a regime of \textit{duality} in
which nuclear matter can be equivalently described in terms of quarks
and diquarks (i.e., the McLerran-Pisarski conjecture as advocated by
\citet{Fukushima:2013rx}).  Such a special regime in the
intermediate baryon density is called the quarkyonic regime.
\end{itemize}

In this work we adopt this last picture of the quarkyonic scenario.
There are three reasons why we consider that the quarkyonic scenario
should be the most realistic:

\noindent
(1) Center symmetry is more and more badly broken with increasing
baryon density, and so the deconfinement phenomenon at high baryon
density should be even broader crossover than that at high
temperature.  Thus, a first-order phase transition to quark matter is
quite unlikely.  In many models the chiral phase transition at low
temperature could be of first order, but once a reasonable amount of
vector coupling is included, it also becomes smooth crossover
\citep{Kitazawa:2002bc,Sasaki:2006ww,Fukushima:2008is,Bratovic:2012qs}. 
Also, the CSC phase can be smoothly connected to the confined phase;
for example $\mathrm{SU(N)}$ lattice gauge theories
coupled to fixed-length Higgs fields
have a crossover region between confined and Higgs phase \citep{Fradkin:1978dv}.
Therefore, it is the most conceivable that nucleons are gradually 
taken over by more fundamental degrees of freedom.

\noindent
(2) It is highly non-trivial how to reconcile confinement and the
BEC-BCS crossover.  
To form bound states of diquarks, the inter-quark
interaction should be strong, and then, the interaction among any
colored objects including diquarks must be strong as well.  Then,
all the colored objects are expected to form color-singlet bound
states.  Theoretically speaking, however, what we can observe in
principle is only the gauge invariant quantity and so there is no way
to judge whether quarks and diquarks are confined in color-singlet
bound states or not.  This at the same time means that we cannot
exclude a mixture of quarks and diquarks even in the confined phase
where the interaction is strong.

\noindent
(3) The hadron resonance gas model is known to be successful to
reproduce the thermodynamic properties of hadronic matter near or even
slightly above $\Tc$ for small $\muB$ (for a recent review, see \cite{Ding:2015ona}). 
This is a clear example of the
duality between hadrons and quarks in a transitional regime.  From
this point of view we can say that the quarkyonic regime could be a
high-density counterpart of the so-called strongly-correlated QGP
(i.e., sQGP) established in RHIC experiments.  This analogy between
sQGP and quarkyonic matter is addressed by \citet{Fukushima:2014pha}.

\subsection{Quark-hadron continuity}

From the point of view of the symmetry-breaking pattern we can give
some more solid arguments to justify the quarkyonic scenario which is
quite consistent with the CSC theory.  It has been a closely
investigated conjecture called the quark-hadron continuity that the
superfluid nuclear matter is a dual state of the CFL phase
\citep{Schafer:1998ef,Alford:1999pa,Fukushima:2004bj,Hatsuda:2006ps}.
The point is that the superfluid nuclear matter and the CFL phase
break chiral symmetry and $\mathrm{U(1)_B}$ symmetry.  The theoretical
description based on the symmetries is elegant and convincing only in
the chiral limit, but the strange quark mass comparable to $\LQCD$
introduces subtlety.  In particular the CFL phase may not continue to
the normal nuclear density due to the Fermi surface mismatch.  Then,
possibilities include a CFL state entering the gapless region which is
known to suffer instabilities against spatial modulation
\citep{Fukushima:2004zq,Huang:2004bg,Fukushima:2005cm}.  If the
interaction is sufficiently strong, the instability might be avoided
\citep{Gubankova:2006gj}.

In this work we postulate a large enough value of the diquark coupling
$H$, so that we do not have to cope with instability problems.  Then,
there should be a phase transition to the two-flavor super-conducting
(2SC) phase down from the CFL phase as lowering the baryon density.
The pure 2SC phase has diquark condensates of $ru$-$gd$ and $rd$-$gu$
quark pairs in color space of red and green and in flavor space of up
and down.  Within the two-flavor sector these combinations in
flavor space are SU(2) singlets, and so none of
$\mathrm{SU(2)_L\times SU(2)_R}$ is broken.  Besides,
$\mathrm{U(1)_B}$ is broken but a modified $\mathrm{U(1)_{\tilde{B}}}$
is kept unbroken, where $\mathrm{U(1)_{\tilde{B}}}$ is generated by a
mixed charge $\tilde{B}=B-2Q_e$.  In fact, the diquarks $ru$-$gd$ and
$rd$-$gu$ have $B=1/3+1/3=2/3$, the electric charge
$Q_e=2/3-1/3=1/3$ in unit of $e$, and therefore $\tilde{B}=0$.  Hence,
no global symmetry is broken in the 2SC phase.

It is also possible to have not the pure 2SC phase but the coexisting
2SC phase with non-zero chiral condensates.  In this case chiral
symmetry is spontaneously broken by the chiral condensates, and the
diquark condensate breaks no new symmetry.  Thus, the coexisting 2SC
phase has a completely identical pattern of the symmetry breaking as
the hadronic phase.  In other words we have no way to exclude the
2SC-type diquark condensate even in NM in the confined phase.
Considering the strange quark mass, such continuity between the
coexisting 2SC phase and NM should be a more realistic candidate than
the CFL-NM continuity.  In conclusion, the quarkyonic scenario based
on deconfinement is supported from the chiral symmetry point of view
by the presence of the coexisting 2SC phase with non-zero chiral
condensates.

Here, let us make two remarks to clarify possibly confusing points.
The first point is the order parameter and gauge symmetry.  In the
mean-field approximation there can be a phase transition between
normal QM and the 2SC phase and the associated order parameter is the
diquark condensate.  So, one might think that there should be also a
phase transition from NM to the 2SC phase even though the symmetry has
no difference.  This is true only approximately as long as one fixes
the gauge and relies on the mean-field approximation.  However,
generally beyond the mean-field level, there is no gauge-invariant
order parameter for the 2SC phase while one could construct higher
dimensional order parameters for the CFL phase
\citep{Rajagopal:2000wf,Fukushima:2010bq,Fukushima:2004bj}.  This
``non-existence'' of the 2SC order parameter strongly suggests
crossover between NM and the 2SC phase, which is reminiscent of the
non-existence of the exact order parameter and smooth crossover of
deconfinement at high temperature.

The next remark is about the consistency of the quarkyonic regime
that was recognized in the large-$\Nc$ limit and the CSC that is
disfavored in the large-$\Nc$ limit.  This is sometimes a source of
conceptual confusion.  To discuss the quarkyonic regime we do not have
to take the large-$\Nc$ limit; while the quarkyonic picture was initially studied
in the large-$\Nc$ limit, its implication is more general
and the essence can be discussed 
even in two-color QCD \citep{Brauner:2009gu,Hands:2010gd}.
If NM and the 2SC phase are
identifiable, we should consider that this continuity between NM and
the 2SC phase is nothing but a clear realization of the quarkyonic
regime in the real world with three colors.  This point is emphasized
by \citet{Fukushima:2014pha} and is precisely the physics picture that
we make full use of in the present work.

\section{Model and parameters}
\label{sec:model}

In the low density region we adopt the EoS of nuclear matter, and in
this section let us explain how we compute the EoS of QM at high baryon density.
We will treat the NJL model within the mean-field approximation
(for details, see e.g. \cite{Buballa:2003qv,Klahn:2006iw,Kojo:2015fua}). 
We have three gap
energies, $\Delta_{ud}$, $\Delta_{ds}$, and $\Delta_{su}$, associated
with diquark condensates.  Moreover, we should introduce three chiral
condensates, $\langle\bar{u}u\rangle$, $\langle\bar{d}d\rangle$, and
$\langle\bar{s}s\rangle$, which are all dynamically determined from
the gap equations.  As we discuss in details, the vector coupling is
an essential ingredient for our prescription, so we need to treat
$n=\langle \bar{q}\gamma^0 q\rangle$ as a mean-field variable, as well
as three chemical potentials $\mu_e$, $\mu_3$, and $\mu_8$ to impose
electric and color charge neutrality.  Therefore, in total, there are
ten mean-field variables and we must solve ten gap equations
simultaneously.

With these mean-field variables (with a notation;
$\phi_u=\langle\bar{u}u\rangle$, $\phi_d=\langle\bar{d}d\rangle$, and
$\phi_s=\langle\bar{s}s\rangle$) we can write the pressure in the
following way:
\begin{align}
  P = & \frac{1}{8\pi^2}\int_0^\Lambda dp\,p^2 \sum_{i=1}^{72}
  \epsilon_i(p) \notag\\
  & - \gs(\phi_u^2+\phi_d^2+\phi_s^2) - 4\gd\phi_u\phi_d\phi_s
   + \gv n^2 \notag\\
  & - \frac{1}{H}(\Delta_{ud}^2+\Delta_{ds}^2+\Delta_{su}^2)
  + P_e + P_\mu\;,
\end{align}
where $\epsilon_i(p)$ represents the energy dispersion relations
obtained from the Dirac Hamiltonian in the Nambu-Gorkov doubled
basis.  The last terms, $P_e$ and $P_\mu$, represent the pressure
contributions from electrons and muons.

In the presence of the mean-field $n$, the quark chemical potential is
renormalized as $\mu_r = \mu-2\gv n$.  Then, the quark chemical
potential takes a from of $9\times9$ matrix given by
\begin{equation}
  \boldsymbol{\mu} = \mu_r \one_{\rm c}\otimes\one_{\rm f}
  - \mu_e \one_{\rm c}\otimes Q_e + (\mu_3 T^3_{\rm c}
  + \mu_8 T^8_{\rm c})\otimes\one_{\rm f}\;,
\end{equation}
where $Q_e=\diag(2/3,-1/3,-1/3)$ is the charge matrix in flavor
space.  The explicit form of the gap matrix with $\Delta_{ud}$,
$\Delta_{ds}$, and $\Delta_{su}$ can be found in the literature
\citep{Fukushima:2004zq}.

For the model parameters, we use the set of \citet{Hatsuda:1994pi}:
\begin{align}
&\Lambda = 631.4\, {\rm MeV}\,,
~~ \gs \Lambda^2 = 3.67\,,
~~ \gd \Lambda^5 = -9.29 \,,
\notag \\
& m_{u,d} = 5.5\,{\rm MeV}\,,
~~ m_s =135.7\,{\rm MeV}\,,
\end{align}
which are fixed to reproduce hadron phenomenology in the vacuum.  The
other parameters $H$ and $\gv$ are left unfixed, and we will use this
freedom to adjust $H$ and $\gv$ to realize the quark-hadron duality
interpolating the APR and CFL EoS's.

\section{Interpolating the EoS with $\gv$-running 2SC}
\label{sec:EoS}

Because the CFL solution does not exist in the density region near
NM, which indicates a first-order phase transition as discussed later,
a natural candidate for the QM ground state there is the 2SC state.  As
elucidated comprehensively in Sec.~\ref{sec:diquark}, the diquark
condensate $\langle ud\rangle$ in the 2SC phase breaks no new symmetry
and so it can coexist in the hadronic phase.  This means that the 2SC
phase could be smoothly connected to NM and, furthermore, the
quarkyonic scenario requires the presence of duality region where NM
and 2SC can represent the same physical system.

It is still non-trivial how to formulate such duality in a practical
way.  We use the APR hadronic EoS for the baseline, while the 2SC
phase has several unconstrained model parameters such as $H$ and
$\gv$.  Our strategy is to make use of these uncertainties positively
to reproduce APR in the low density region in a spirit of the
quarkyonic scenario.  To this end we treat $\gv$ in the 2SC phase as a
$\muB$-dependent control parameter and fit the resulting EoS with
APR.\ \ Besides, we emphasize that this choice of $\gv$ as a control
parameter would provide us with convenient and transparent intuition
about modified QM EoS, and at the same time, that $\gv$ should be a
\textit{representative} of all unknown effects like confinement near
the normal nuclear density.  The important point is that in this way,
once $\gv(\muB)$ is determined, we can smoothly extend APR toward QM
without any artificial manipulations.

The pressure $P$ in the 2SC phase depends monotonically on $\gv$, so
that we can easily find $\gv$ that reproduces APR at each $\muB$.  One
might think that a $\muB$-dependent $\gv$ may change one of the gap
equations, $\partial P/\partial n=0$.  For technical simplicity we
do not change the gap equation but treat $n$ as an internal
mean-field variable which takes a different value from $\nB$.  Thus,
to obtain $\nB$ correctly, we have to compute
$\partial P/\partial \muB$ including the derivative on $\gv(\muB)$.

\begin{figure}
 \includegraphics[width=\columnwidth]{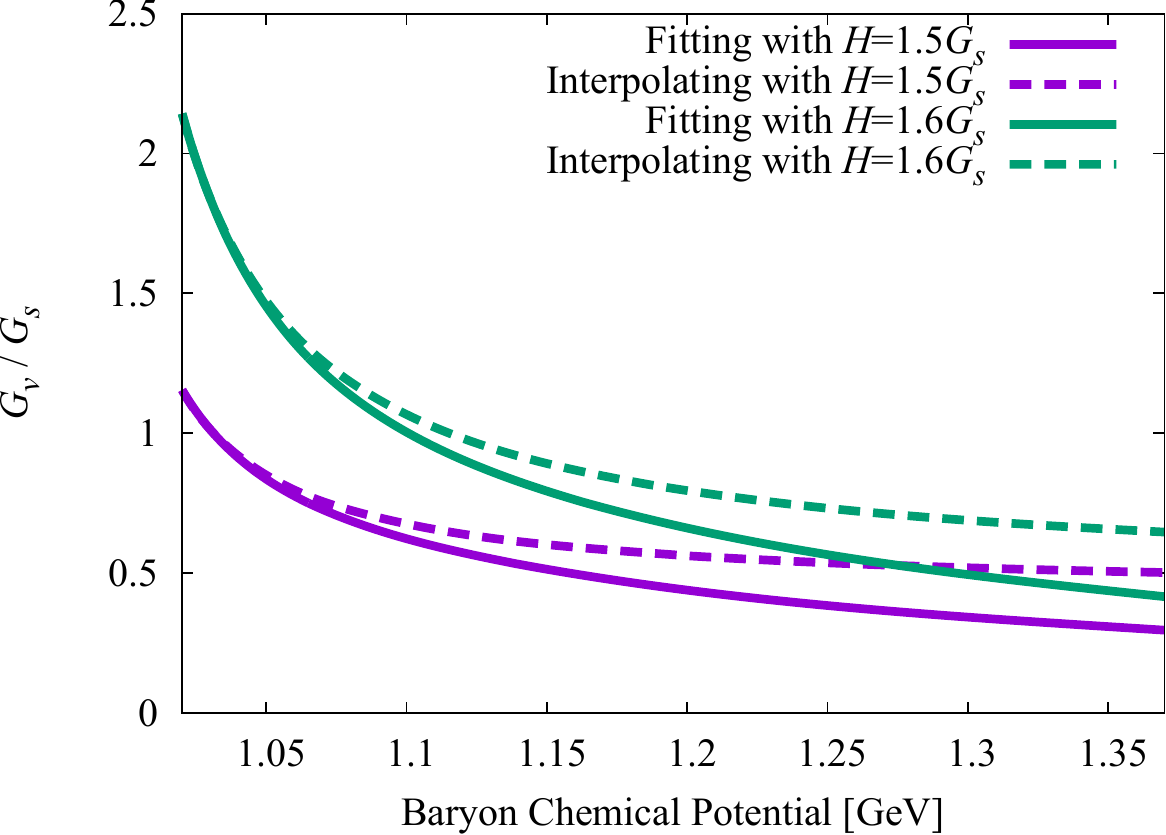}
 \caption{Vector coupling in the 2SC phase fitted with APR for
   $H=1.5\gs$ (lower solid curve) and $H=1.6\gs$ (upper solid curve).
   The interpolating fit results to the CFL phase with $d=0.4$ (See Eq.\ref{eq:gv}) 
   are
   represented by the dotted curves.}
 \label{fig:gv}
\end{figure}

\begin{figure}
 \includegraphics[width=\columnwidth]{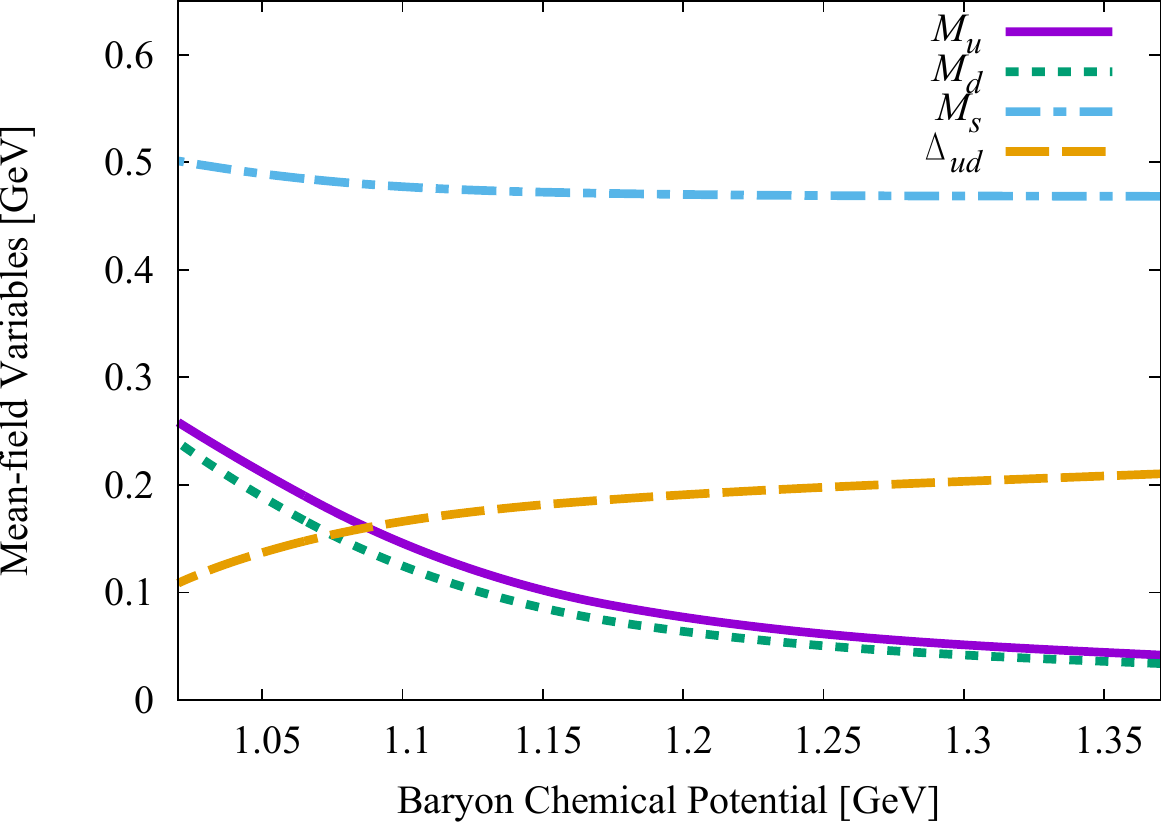}
 \caption{Constituent quark masses and the diquark gap $\Delta_{ud}$
   in the 2SC phase as functions of $\muB$ in the case of $H=1.5\gv$
   (corresponding to the solution shown by a lower solid curve in
   Fig.~\ref{fig:gv}).  The chiral phase transition is a smooth
   crossover because of the vector coupling.}
 \label{fig:cond}
\end{figure}

In order to find a suitable running coupling, we found it necessary to take
a diquark coupling larger than $1$. Otherwise the NJL pressure at any positive $\gv$ 
undershoots APR at typical $\mu_B$, because the NJL effective quark mass, $M_q \simeq 336\MeV$, 
allows the pressure to develop only when $\mu_B > 3M_q \simeq 1010\MeV (> m_N)$ ($m_N$: nucleon mass).
This situation is cured by including the sufficiently large diquark coupling which
effectively reduces the effective quark mass \citep{Kojo:2014rca}. 
Similar effects can be seen in constituent quark models where the color hyper-fine interaction
makes the nucleon mass  smaller than three times constituent quark mass.
We will be interested in $\gv \gtrsim 0.5\gs$ to achieve sufficient stiffness, 
and then $H\gtrsim 1.4\gs$ is required to pass the $2M_\sun$ and causality constraints.
Therefore we will examine  $H=1.5\gs$ and  $H=1.6\gs$ in the following.

In our fitting procedures one of questions is how to choose the fitting range. 
Let us first consider a relatively wide fitting range, $\muB=(1.02\sim 1.20)\GeV$
or $n_B \simeq(1.5 \sim 3.5) n_0$.
(We omitted a very dilute region $\muB\lesssim 1.02\GeV$ because it requires too much fine-tuning in our fit functions.)
We show our numerical results with running $\gv$ for $H=1.5\gs$ (lower
solid curve) and $H=1.6\gs$ (upper solid curve) in Fig.~\ref{fig:gv}.
We emphasize that we did not assume any functional form
\textit{a priori} and $\gv(\muB)$ shown in Fig.~\ref{fig:gv} results
solely from the fit to APR once we make a choice of the diquark
coupling $H$.  We vary $H$ to check the sensitivity and will see that
this choice is near the upper limit not to violate the causality.  
It is important to note that the chiral phase transition is a very smooth
crossover in the presence of large $\gv$, so that this 2SC phase can
accommodate both diquark and chiral condensates for any $\muB$.  This
is clear in Fig.~\ref{fig:cond} where the constituent quark masses,
$M_u$, $M_d$, $M_s$, and the gap energy $\Delta_{ud}$ are given as
functions of $\muB$.  There is a small discrepancy between $M_u$ and
$M_d$ because of the electric charge neutrality condition that breaks
isospin symmetry.

Our fit to APR tells us that $\gv$ must be very large at low $\mu_B$ and relax to $\gv\sim 0.5\gs$ at large $\mu_B$.
The importance of $\gv$ can be inferred from nuclear physics (e.g. $\omega$-meson exchange) 
and typical magnitude $\gv\sim 0.5\gs$ in our fit is understandable within the conventional context.
On the other hand, very large value at low $\mu_B$ represents something else and requires explanations.
Let us recall that using the diquark coupling 
we were able to make the onset chemical potential of the NJL pressure close to APR, 
as required from the quark-hadron duality.
But there remains another problem in our fitting procedures; 
pressure and number density grow much faster in the NJL model than in APR.
The obvious reason is the lack of confinement in our model;
a quark excites individually, not as a part of a baryon, producing artificial excess of pressure.
But large $\gv$ can be used to imitate confining effects by tempering the growth of quark number density.
This effect should be stronger at lower density;
as baryons overlap, $\gv$ becomes less responsible for confining effects
and should relax to the value expected from the repulsive nuclear forces.

Interestingly, we have found that the best fit form of $\gv(\muB)$ is
an inverse logarithm for both $H=1.5\gs$ and $H=1.6\gs$.  Such an
inverse logarithmic is quite suggestive because it is consistent with
the common form of the running coupling constant at one-loop level.
However, the validity of this fitting should be lost at some point of the baryon density. 
In fact, at sufficiently high baryon density there is a phase transition 
from the 2SC to CFL phase which softens the EoS.
If we use the running $\gv(\muB)$ determined above,
the EoS including the CFL phase does not support the massive
neutron star with $M\gtrsim 2M_{\odot}$;
$\gv$ should be $\sim 0.5\gs$ or greater in the CFL phase.  
Thus we need stiffer 2SC pressure with which the
EoS can remain stiff enough even after the transition to the CFL phase occurs.

The above determination used fitting over the range $n_B \simeq(1.5 \sim 3.5) n_0$,
but below we shall relax the condition and use only the range $n_B \simeq (1.5\sim 2)n_0$,
which corresponds to $\muq=(340\sim 345)\MeV$ (i.e., $\muB=(1.02\sim 1.035)\GeV$).
Then the 2SC pressure can be stiffer than before in the range of  $n_B \simeq (2\sim 3.5)n_0$.
It is reasonable to stop fitting at $\muB=1.035 \GeV$, because the baryon density is $n_B \simeq 2n_0$
beyond which APR is no longer reliable.

To satisfy the boundary conditions, i.e., the smooth connection to APR in the lower-density
side and to the CFL phase with $\gv\gtrsim 0.5\gs$ in the higher-density side, 
we modify $\gv(\muB)$ from an inverse
logarithm to the following form:
\begin{equation}
 \gv(\muB)/\gs = \frac{a}{\log[(\muB-b)/c]} + d
\label{eq:gv}
\end{equation}
with an offset by $d$.  Once we fix $d$, we can determine other three
parameters, $a$, $b$, $c$ using the smooth connection to APR.\ \ We
changed $d$ to find that the massive neutron star with $M>2M_{\odot}$
is impossible with $d\lesssim 0.3$.  We shall therefore choose $d=0.4$
throughout this work.  The parameters fixed in such a way, for
$H=1.5\gs$ and $1.6\gs$ respectively, are listed in
Tab.~\ref{tab:param} and the corresponding $\gv(\muB)$ that
interpolates between APR and the CFL phase is overlaid by dashed
curves in Fig.~\ref{fig:gv}.  

\begin{table}
 \begin{tabular}{cc|ccc}
 \hline
 $H/\gs$ & $d$ & $a$ & $b$ [GeV] & $c$ [GeV] \\
 \hline
 1.5 & 0.4 & 0.05283 & 0.4049 & 0.5735 \\
 1.6 & 0.4 & 0.1127 & 0.2942 & 0.6804 \\   
 \hline
 \end{tabular}
 \caption{Parameters for the interpolating $\gv(\muB)$ between APR and
   the CFL phase.}
 \label{tab:param}
\end{table}

\begin{figure}
 \includegraphics[width=\columnwidth]{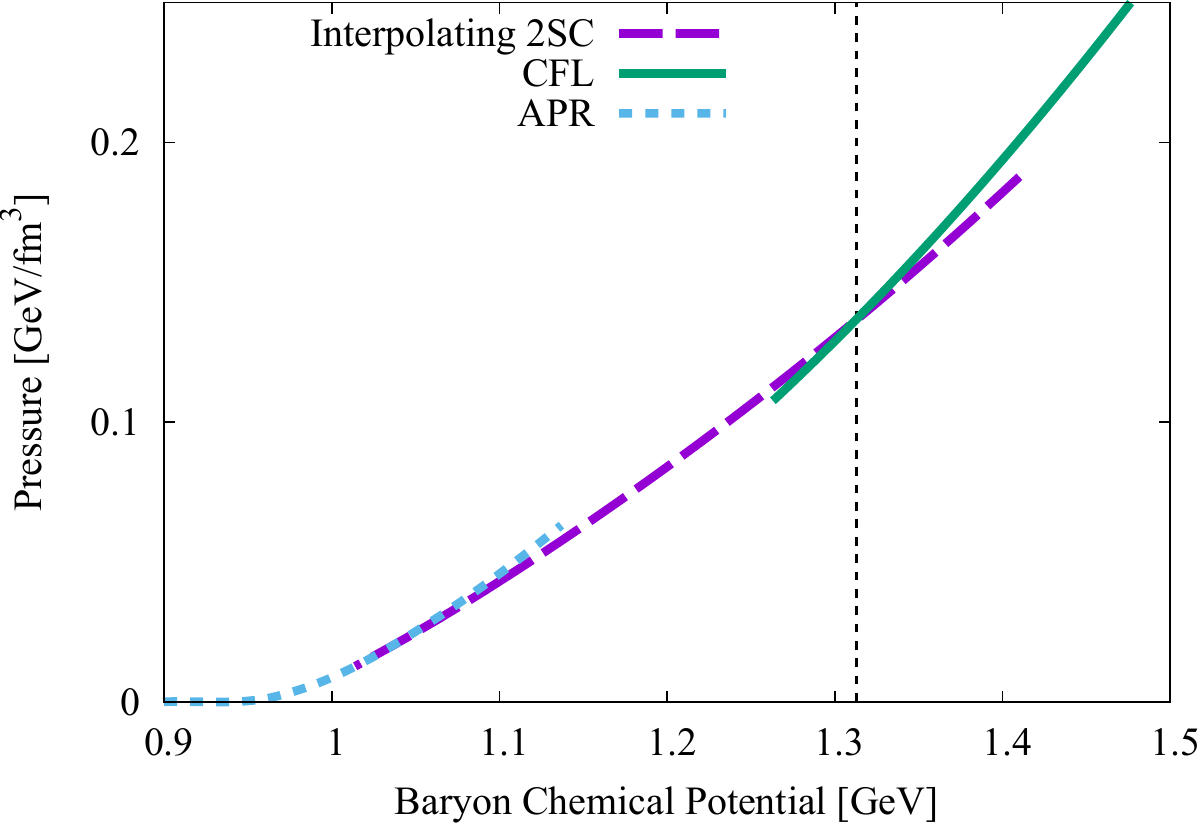}
 \caption{Pressure comparison of the 2SC phase (dashed curve) and the
   CFL phase (solid curve) for $H=1.5\gs$.  The pressure of APR
   (dotted curve) is also shown for reference.}
 \label{fig:pressure}
\end{figure}

With this running-$\gv$ we can find the CFL solution as well as the
2SC phase and then we can locate a first-order phase transition
between them by comparing the pressure.  In Fig.~\ref{fig:pressure} we
show an example for $H=1.5\gv$ to find a first-order phase transition
at $\muB=1.31\GeV$ where the pressure of the interpolating 2SC phase
with running $\gv$ and the CFL phase crosses.  We make a remark on the
connection between APR and the 2SC phase around $\muB\sim 1\GeV$.
From Fig.~\ref{fig:pressure} one might think that APR has a slightly
larger pressure above the fitting region, and so APR would be rather favored.  
To resolve such confusion we here again emphasize our
picture of the quarkyonic scenario.  The change from NM to QM is not
any phase transition but what we assume is a dual regime around
$\muB\sim 1\GeV$ in which NM is gradually taken over by QM. 
\ \ In contrast to this smooth crossover from NM to QM, the change from the
2SC phase to the CFL phase is a genuine physical phase transition
with different symmetry-breaking patterns.  In many model studies
including the present work, this phase transition turns out to be of
first order.

\begin{figure}
 \includegraphics[width=\columnwidth]{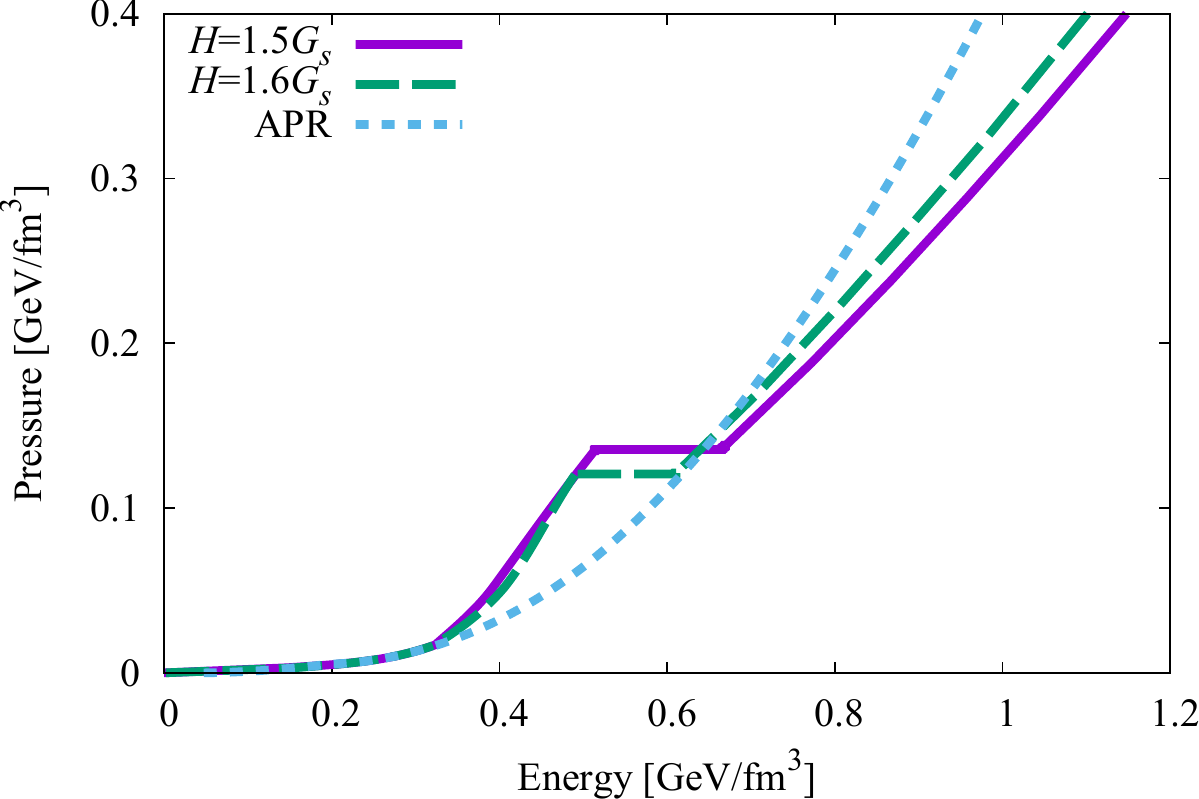}
 \caption{Pressure $P$ as a function of the energy $\varepsilon$ for
   $H=1.5\gs$ (solid curve) and $H=1.6\gs$ (dashed curve).  The
   pressure of APR extrapolation (dotted curve) is shown for
   reference, though the plotted range is outside of its validity
   region.}
 \label{fig:eos}
\end{figure}

Now that we have the EoS for the whole range of $\muB$ from NM to CSC,
we can compute not only the pressure $P$ but the energy
$\varepsilon=\muB \nB-P$ as well.  Actually, the relation of $P$ vs
$\varepsilon$ is essential for the estimation of the neutron star
mass.  Because $\varepsilon$ involves a first derivative in $\nB$, its
value jumps discontinuously at the first-order phase transition.  We
can see this behavior in our numerical results shown in
Fig.~\ref{fig:eos}.  It is also clear in Fig.~\ref{fig:eos} that the
2SC part hardly changes with different choices of $H$.  We can explain
this minor dependence from the fact that we impose the same boundary
condition of APR at lower density.  
The other boundary condition of
the CFL phase side is, on the other hand, 
loosely constrained by APR, and so there remains $H$ dependence in the CFL
part as is the case in Fig.~\ref{fig:eos}.  
But as we will see below, the value of $H$ is constrained by the causality condition.

\begin{figure}
 \includegraphics[width=\columnwidth]{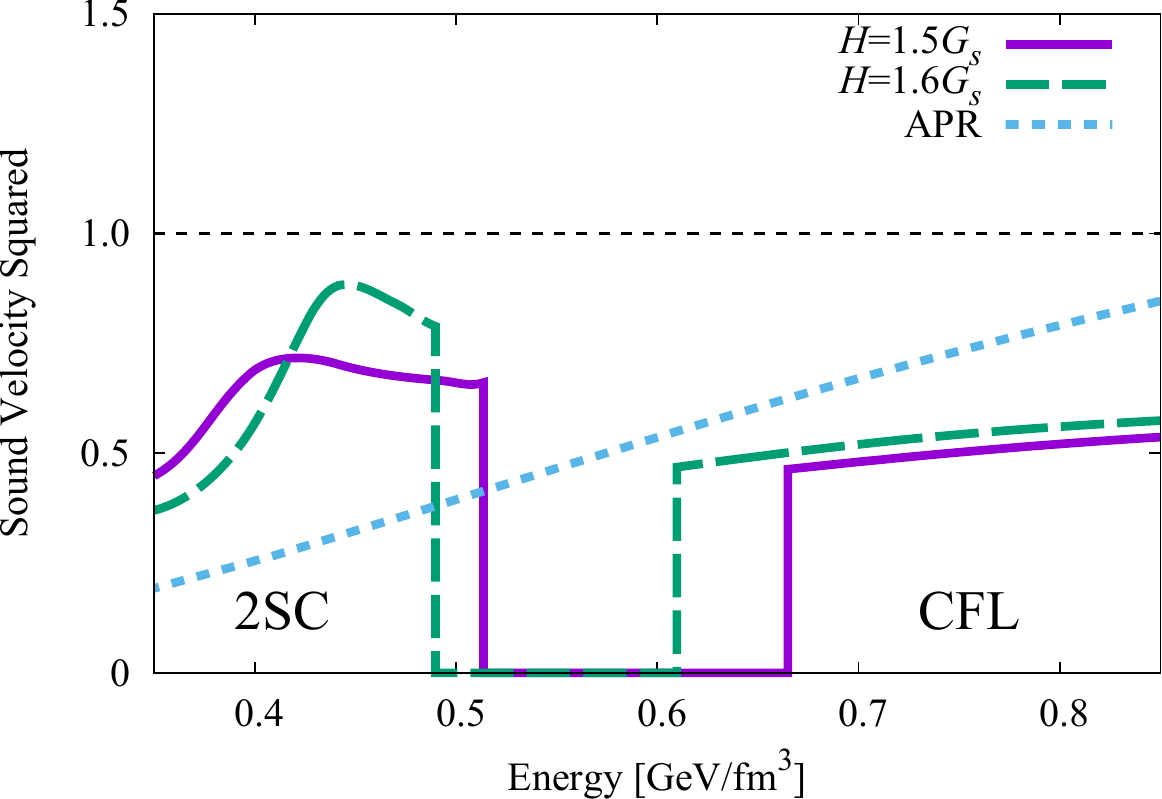}
 \caption{Sound velocity squared as a function of the energy for
   $H=1.5\gs$ (solid curve) and $H=1.6\gs$ (dashed curve).  The APR
   extrapolation (dotted curve) is shown for reference.}
 \label{fig:sv}
\end{figure}

One can deduce the sound velocity (squared)
$c_s^2 = \partial P/\partial \varepsilon$ from the slope of curves in
Fig.~\ref{fig:eos}.  We can numerically take the derivative of $P$
with respect to $\varepsilon$ and present obtained $c_s^2$ in
Fig.~\ref{fig:sv}.  It is important to check the causal condition that
$c_s^2$ should not exceed the unity (i.e., the speed of light in the
natural unit).  In the present setup, as understood from
Fig.~\ref{fig:sv}, the causality is not violated.

The structure of $c_s^2$ is almost the same irrespective of the choice
of $H$;  with increasing $\varepsilon$ it monotonically increases in
the APR region and then has a peak in the 2SC phase.  Once a
first-order phase transition to the CFL phase occurs, $c_s^2$ is
pushed down to $\sim 0.5$.  In the limit of large $\varepsilon$ or
high baryon density, $c_s^2$ asymptotically approaches $\sim 0.6$
which slightly depends on $H$.  An interesting observation in
Fig.~\ref{fig:sv} is that the peak height strongly depends on $H$ and
if $H$ is greater than $\sim 1.6\gs$, it would go beyond the unity,
which would violate the causality.  Therefore, such a large $H$ is not
allowed and $H=1.6\gs$ is close to the upper limit for our choice of
$d=0.4$ in Eq.~\eqref{eq:gv}.  In this way we find that there is not
much uncertainty in the EoS determination after all.

Finally we comment on the connection of our result to the pQCD EoS. 
If we extrapolate our $P(\mu_B)$ curve to $\mu_B \sim 3\GeV$,
it undershoots the pQCD pressure.
This is not quite surprising: 
in the pQCD domain, one should expect smaller $\gs$, $\gv$, and $H$
than those used in our analyses, since non-perturbative effects (gluons)
are irrelevant by definition of the pQCD domain. 
By simply implementing the reduction of NJL parameters at large $\mu_B$,
one can connect our EoS to the pQCD EoS at $\mu_B \gtrsim 3\GeV$.

\section{Mass-radius relation}
\label{sec:MR}

To solve the $M$-$R$ relation from the TOV equation what we need is
the EoS shown in Fig.~\ref{fig:eos}.  Plugging our EoS to the TOV
equation and changing the initial condition that is the central
pressure at $r=0$, we can get a curve in the plane of the mass and the
radius of the neutron star as is just the standard procedure.

\begin{figure}
 \includegraphics[width=\columnwidth]{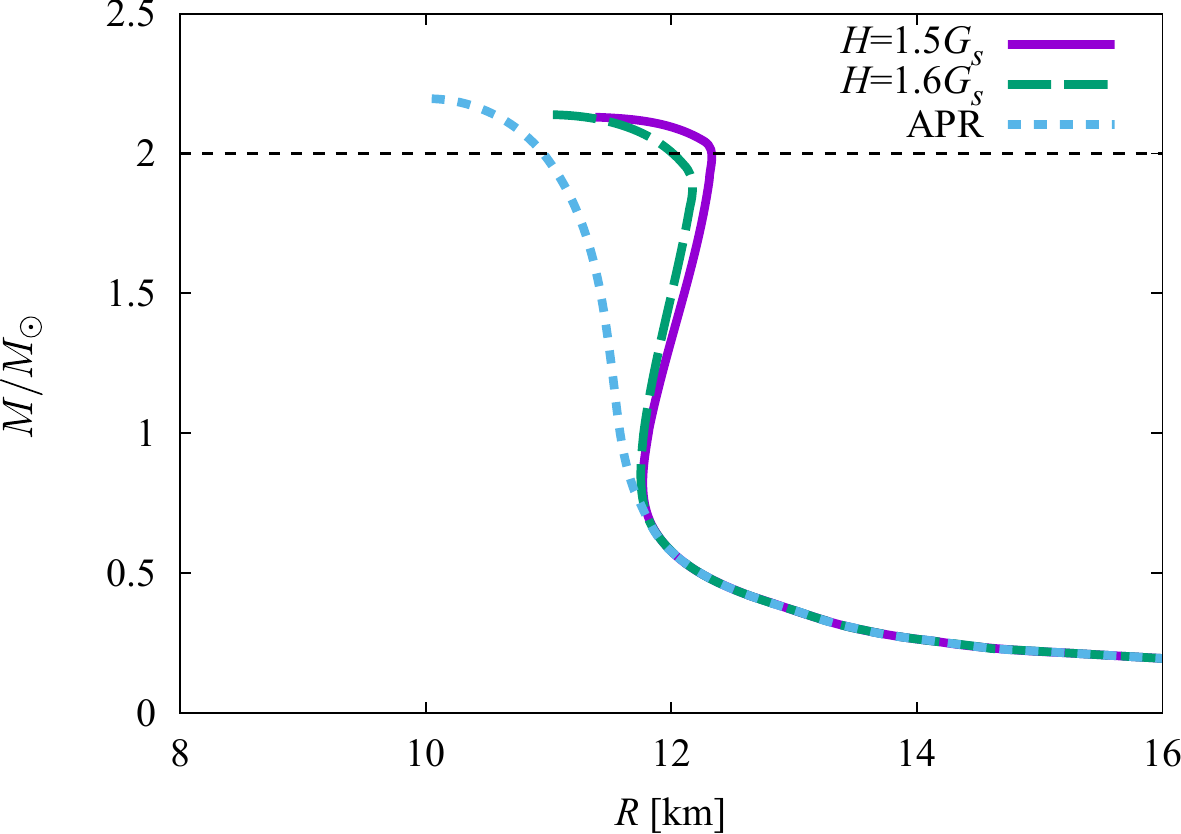}
 \caption{$M$-$R$ relation for $H=1.5\gs$ (solid curve) and $H=1.6\gs$
   (dashed curve).  The APR result (dotted curve) is shown for
   reference.}
 \label{fig:m_r}
\end{figure}

We summarize our results in Fig.~\ref{fig:m_r}.  The $M$-$R$ curves
below $M_\odot$ are essentially determined by the APR EoS up to
$\nB\sim 2n_0$.  (The tail at large $R$ is 
determined by the crust EoS, for which we adopt the SLy model by
\citet{Douchin:2001sv}.)  Then, the 2SC curves start to take off from
the APR by degrees.  The 2SC curve's having the larger radii than the
APR originates from a stiffer EoS in the 2SC phase at $\nB>2n_0$.  In
the vicinity of $M\sim 2M_\odot$ the curves have a turning point which
signals the phase transition from the 2SC phase to the CFL phase.  The
maximal mass in our parametrization reaches $\simeq 2.2 M_\odot$.

The neutron star radii at the canonical mass $1.4M_\odot$ typically
spread over $R=(9\sim 16)$ km, depending on the EoS for asymmetric
nuclear matter.  The radius of $R\simeq 12$ km in our results belongs
to a group of small radii, 
reflecting that the APR at low density is softer than other typical hadronic EoS 
such as mean-field EoS.
The small radii are consistent with recent observation that favors $R=(10\sim 13)$ km
\citep{Ozel:2010fw,Ozel:2015fia,Steiner:2010fz,Steiner:2012xt},
though those analyses still contain several assumptions to be verified.
On the other hand there is also a group insisting 
$R> 14$ km \citep{Suleimanov:2010th},
which favors stiff hadronic EoS at low density
(for hybrid EoS with stiff hadronic EoS, see e.g. 
\cite{Bonanno:2011ch,Benic:2014jia}).
Another hint for the small radii comes from recent quantum Monte-Carlo
calculations which indicate that nuclear EoS at low density is
relatively soft, or even slightly softer than the APR
\citep{Gandolfi:2011xu}.  The heavy-ion data by
\citet{Danielewicz:2002pu} also favor a soft EoS at low density.

Our crossover EoS yields relatively simple topology for the $M$-$R$ curves
similar to those in purely hadronic EoS.
In case of the hybrid EoS with a first order phase transition,
there are more varieties of the $M$-$R$ curves.
The classification was done by \cite{Alford:2013aca,Alford:2015gna},
assuming the constant speed of sound.

\section{Conclusions}
\label{sec:conclusions}

Quarkyonic matter is a likely candidate to bridge a gap between
nuclear matter and quark matter.  It is an intermediate state of
matter with a duality region that is describable by baryons or
quarks.  Thus, the EoS construction is not a patchwork of distinct
EoS's of nuclear and quark matter, as is a conventional hybrid
description having first-order phase transitions.

We discussed the interplay among quarks, diquarks, and baryons.
Through processes of exchanging mesons or quarks between baryons, the
diquark correlation develops in the intermediate state.  While the
density is low, the diquark is combined with other colored excitations
to form local and color-singlet objects.  For this reason diquarks
should become manifest increasingly as baryons involve more and more
quark exchange.  We classified scenarios of deconfinement crossover
according to the diquark characters and addressed what quarkyonic
matter implies.

To interpret the crossover EoS in the language of microscopic degrees
of freedom, we introduced a running vector coupling for quark models,
with which the quark-hadron duality is implemented in a practically
feasible manner.  The running coupling is constrained by the
conventional nuclear EoS in the low-density side and by massive
neutron stars with $M>2M_\odot$ in the high-density side.  Because the
idea is quite generic, one can employ any quark models and we adopted
an NJL-type model for a concrete demonstration in this work.

The important finding is that the model parameters around
$\nB\sim 10n_0$ must be comparable with those in the vacuum (i.e.\ matter at $\sim 10n_0$ still belongs to a non-perturbative regime);
otherwise, the $2M_\odot$ constraint is not satisfied.  The upper
bound of the diquark interaction comes from the causality condition.
After all, there is not much uncertainty remaining in the parameter
determination.  In principle, all the interactions in the quark model
are mediated by gluons, and so substantially large couplings require
non-perturbative gluon dynamics.  We need to clarify the origin of
non-perturbative dynamics in the high-density limit in a similar sense
to the high-temperature limit at which the magnetic sector is still
confining.

In this paper we interpret the enhancement of many-quark (meson) exchanges at $n_B \gtrsim 2n_0$ 
as an indicator for a new state of matter. 
While we consider diquarks in this paper, another branch of our picture
would be the condensation of mesonic objects since many mesons are available among baryons.
Although the previous studies in hadronic models typically predicted 
strong first order phase transitions associated with meson condensed phases,
recent studies in quark models
indicate that the appearance of inhomogeneous mesonic condensates does not necessarily 
accompany strong first order phase transitions, 
thus can avoid the significant softening \citep{Buballa:2015awa}. 
The Skyrme crystal descriptions at {\it high} density \citep{Paeng:2015noa} have some overlap with
such quark descriptions since both of them can address several effects beyond
purely hadronic descriptions, such as the structural change of hadrons. 

In this work we focused on the bulk properties of matter, i.e., the
EoS only.  To explore the refined characters of matter and to promote
the advantage of our proposed method, we should study transport
processes such as the cooling problem involving the strangeness.  We
shall leave this for future works.

\acknowledgments
We thank G. Baym, K. Masuda, L. McLerran, and J. Wambach 
for insightful conversations during Quark Matter 2015 in Kobe.
This work was partially supported by JSPS KAKENHI Grants
No.\ 15H03652 and 15K13479 (K.\ F.) and by NSF Grants PHY09- 69790 and
PHY13-05891 (T.\ K.).

\bibliographystyle{apj}
\bibliography{quarkyonic}
\end{document}